\documentclass[pra,showpacs,twocolumn,aps]{revtex4}

\usepackage[utf8]{inputenc}
\usepackage{amsmath}
\usepackage{amssymb}
\usepackage{latexsym}
\usepackage{amsfonts}
\usepackage{epsfig}
\usepackage{psfrag}
\usepackage{graphicx}
\usepackage{subfigure}

\usepackage{color}

\newcommand{\ket}[1]{\left|#1\right>}
\newcommand{\bra}[1]{\left<#1\right|}

\newcommand{\f}[1]{\mbox{\boldmath$#1$}}
\newcommand{\fk}[1]{\mbox{\boldmath$\scriptstyle#1$}}

\newcommand{\bea}{\begin{eqnarray}}
\newcommand{\ea}{\end{eqnarray}}
\newcommand{\eea}{\end{eqnarray}}
\newcommand{\ord}{\,{\cal O}}

\begin{document}

\title{On the feasibility of a nuclear exciton laser}

\author{Nicolai ten Brinke}
\author{Ralf Sch\"utzhold}
\email[e-mail:\,]{ralf.schuetzhold@uni-due.de}
\affiliation{Fakult\"at f\"ur Physik, Universit\"at Duisburg-Essen, 
Lotharstrasse 1, D-47057 Duisburg, Germany}

\author{Dietrich Habs}
\affiliation{Fakult\"at f\"ur Physik, Ludwig-Maximilians-Universit\"at 
M\"unchen, Am Coulombwall 1, D-85748 Garching, Germany}

\date{\today}

\begin{abstract}
Nuclear excitons known from M\"ossbauer spectroscopy describe coherent 
excitations of a large number of nuclei -- analogous to Dicke states 
(or Dicke super-radiance) in quantum optics. 
In this paper, we study the possibility of constructing a laser 
based on these coherent excitations. 
In contrast to the free electron laser (in its usual design), 
such a device would be based on stimulated emission and thus might 
offer certain advantages, e.g., regarding energy-momentum accuracy.
Unfortunately, inserting realistic parameters, the window of 
operability is probably not open (yet) to present-day technology -- 
but our design should be feasible in the UV regime, for example.
\end{abstract}


\maketitle

\section{Introduction}
The invention of the laser lead to a giant leap in the field of classical 
and quantum optics. 
This light source offers unprecedented possibilities regarding features 
such as coherence, intensity, and brilliance etc.
Unfortunately, however, it is not easy to transfer this successful concept 
beyond the optical or near-optical regime, 
cf.~\cite{Baldwin:1997ve,Tkalya:2011dq}.
Free-electron lasers, for example, work at much higher energies -- but their 
principle of operation (in their usual design) is more similar to classical 
emission instead of stimulated emission.
As a result, their properties (e.g., regarding coherence) are not quite 
comparable to optical lasers.  

There is another phenomenon in this energy range $\ord(\rm keV)$ in which 
coherence plays a crucial role -- nuclear excitons known from M\"ossbauer 
spectroscopy \cite{Mossbauer:1958fk,Mossbauer:1958kx}.
These coherent excitations of a large number of nuclei 
\cite{Hannon:1999fk,Smirnov:2005uq,Habs:2009uq} are analogous to 
Dicke states \cite{Dicke:1954kx} (also known as Dicke super-radiance 
\cite{Scully:2006fk,Scully:2007fk,Sete:2010fu}) in quantum optics.
The coherence results in constructive interference of the emission 
amplitudes from many nuclei \cite{Burnham:1969uq} and is facilitated 
by the fact that the 
photon recoil is absorbed by the whole lattice 
\cite{Mossbauer:1958fk,Mossbauer:1958kx} instead of the individual 
nuclei (which would destroy the coherence).
For example, the coherent nature of the propagation of nuclear 
excitons through resonant media, showing quantum beats, was observed in 
\cite{Frohlich:1991kx,Burck:1999fk}.
Other cooperative effects of coherently excited nuclei have been studied, 
such as the collective Lamb shift \cite{Rohlsberger:2010fk}, 
coherent control of nuclear x-ray pumping \cite{Palffy:2011vn}, and 
electromagnetically induced transparency \cite{Rohlsberger:2012uq}.

In the following, we study the possibility of constructing a laser-type 
device employing these nuclear excitons, 
which is based on stimulated emission \cite{ELIwhitebook}.
Such a device could combine the advantages of the free-electron laser 
with the coherence and brilliance of nuclear excitons. 

\section{Hamiltonian}
First, we describe a single nucleus as a two-level system with transition 
frequency $\omega$ interacting resonantly with a single-mode field. 
In rotating-wave and dipole approximation, the Hamiltonian can then 
be cast into the standard form ($\hbar = c = \varepsilon_0 = 1$)
\bea
\label{eq:hsingle}
\hat{H}_{\rm single} 
= 
\left( 
g \hat{a} \sigma_\ell^+ e^{i \fk{\kappa} \cdot \fk{r}_\ell} + {\rm H.c.} 
\right) 
+ \frac{\omega}{2} \left( \hat{\sigma}^z_\ell + 1 \right) 
+ \omega \hat{a}^\dagger \hat{a}
\,.
\ea
As usual, the ladder operators 
$\sigma_\ell^{\pm} = ( \sigma_\ell^x \pm i \sigma_\ell^y ) / 2$ 
and the Pauli matrix $\sigma_\ell^z$ describe the two-level system. 
The first term governs the interaction (with coupling constant~$g$)
with the electromagnetic field and 
thus contains photonic annihilation and creation operators 
$\hat{a} / \hat{a}^\dagger$ and phase factors 
$e^{i \fk{\kappa} \cdot \fk{r}_\ell}$ depending on the location 
of the nucleus, $\f{r}_\ell$, and the wavenumber $\f{k}=\f{\kappa}$ 
of the photon mode with $|\f{\kappa}| = \omega$. 
The second and third term account for the energy stored in the two-level 
nucleus and in the single-mode field, respectively.
When dealing with many $S \gg 1$ two-level nuclei instead of one, 
we can sum up the individual-nucleus Hamiltonians and arrive at
\bea
\label{eq:hmany}
\hat{H} 
=
\left( g \hat{a} \hat{\Sigma}^+ + {\rm H.c.} \right) 
+ \omega \left( \hat{\Sigma}^z + \frac{S}{2} \right) 
+ \omega \hat{a}^\dagger \hat{a}
\,,
\ea
where quasispin-$S$-operators have been introduced
\bea
\hat{\Sigma}^\pm 
= 
\sum_{\ell=1}^S \sigma_\ell^\pm \exp\{\pm i \f{\kappa} \cdot \f{r}_\ell\}
\,,\quad
\hat{\Sigma}^z = \frac{1}{2} \sum_{\ell=1}^S \sigma_\ell^z
\,.
\ea
In the interaction picture, the perturbation Hamiltonian, 
originating from the first term in Eq.~(\ref{eq:hmany}), reads
\bea
\label{eq:hinteract}
\hat{V} = g \hat{a} \hat{\Sigma}^+ + {\rm H.c.}
\,.
\ea
The quasispin-$S$-operators 
$\hat{\Sigma}^\pm=\hat{\Sigma}^x\pm i\hat{\Sigma}^y$
and 
$\hat{\Sigma}^z$ generate an $SU(2)$ algebra \cite{Lipkin:2002fk}.
Thus, the transition matrix elements for collective transitions not 
only depend on the number of nuclei involved, but also on the number 
of excitations $s$
\bea
\label{eq:matrix_elements}
\hat{\Sigma}^+\ket{s}&=&\sqrt{(S-s)(s+1)}\ket{s+1}
\,,\nonumber\\
\hat{\Sigma}^-\ket{s}&=&\sqrt{(S-s+1)s}\ket{s-1}
\,,
\ea
where $\ket{s} \propto (\hat{\Sigma}^+)^s \ket{0}$ denotes a coherent 
state with $s$ excitons, often referred to as Dicke states 
\cite{Dicke:1954kx}. 

\section{Coherent emission}
\begin{figure}[h]
\begin{center}
\psfrag{gamma}{$\f{\gamma}$}
\psfrag{kvector}{$\f{k}$}
\psfrag{0}{\scriptsize{$0$}}
\psfrag{1}{\scriptsize{$1$}}
\psfrag{2}{\scriptsize{$2$}}
\psfrag{3}{\scriptsize{$3$}}
\psfrag{S}{\scriptsize{$S$}}
\subfigure[]{\includegraphics[width=0.5\columnwidth]{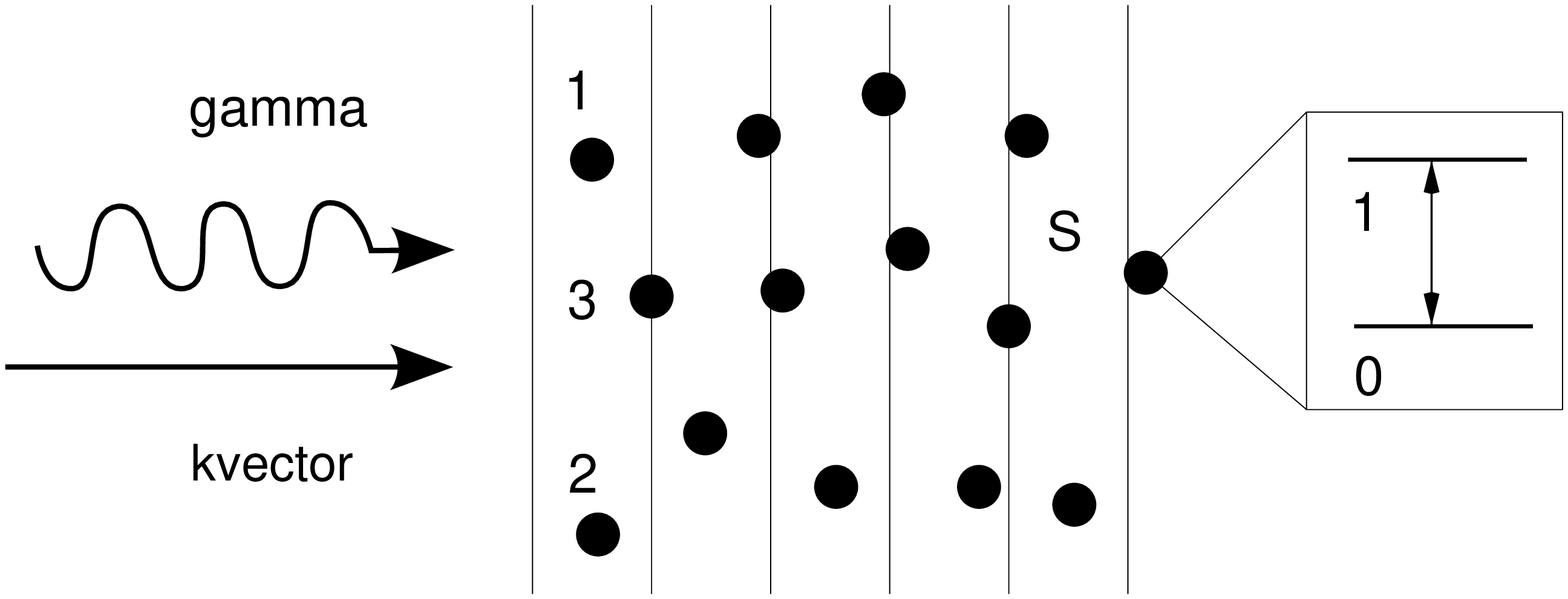}}
\hspace{.5cm}
\subfigure[]{\includegraphics[width=0.4\columnwidth]{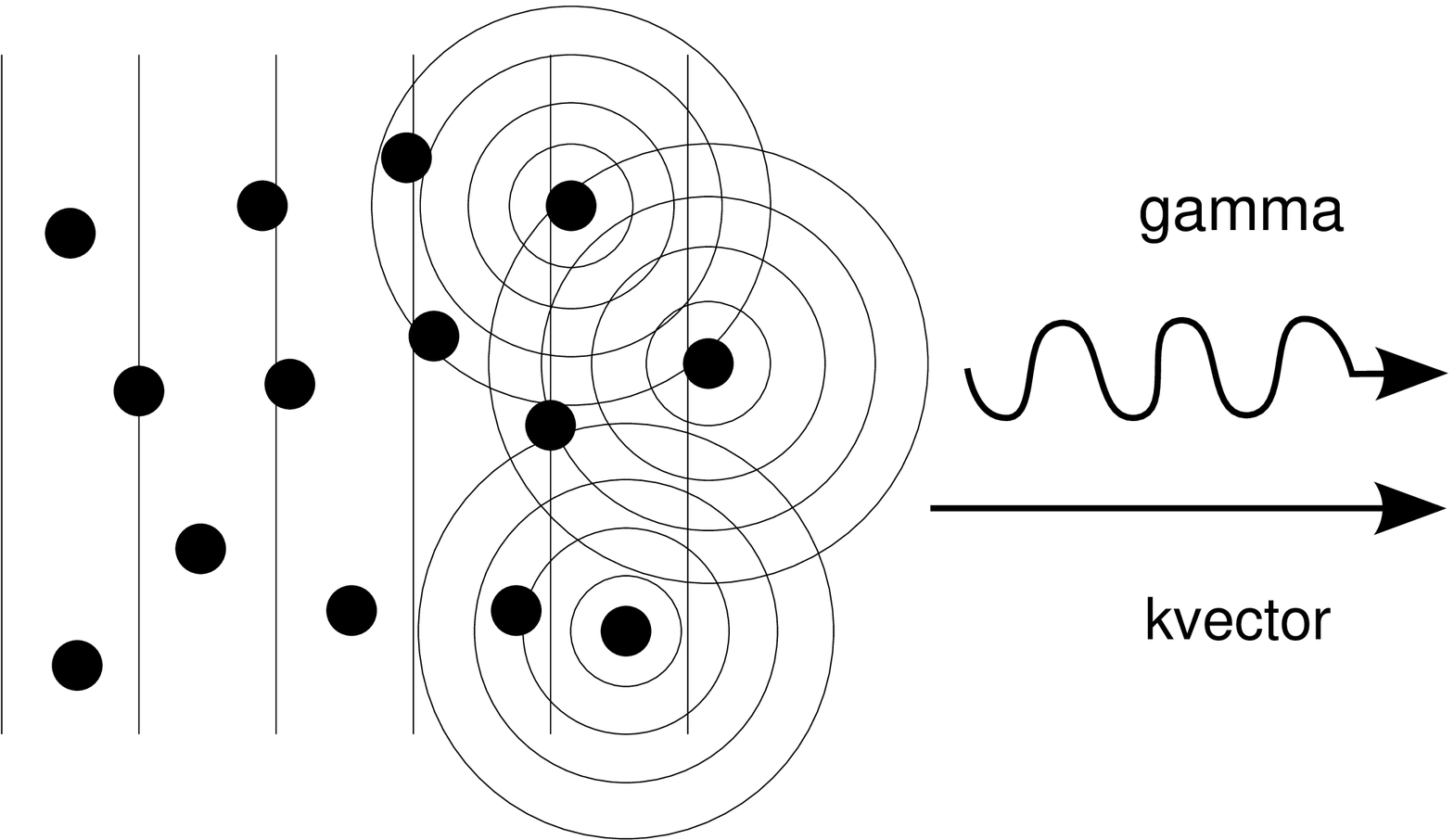}}
\caption{Sketch of the coherent properties of nuclear excitons. 
An incident photon with wave-vector $\f{k}$ is absorbed (a)
by an ensemble of $S\gg1$ nuclei (two-level systems) and thus 
generates a Dicke state $\ket{s=1}$. 
Then the decay amplitudes of all these nuclei add up coherently 
in forward direction such that the absorption is followed by 
collective spontaneous emission into the same direction (b).} 
\label{fig:exciton_and_dicke}
\end{center}
\end{figure}

In contrast to the spontaneous decay of a single excited nucleus, 
where the resulting photon can be emitted in all directions, exciton 
states as in Eq.~(\ref{eq:matrix_elements}) predominantly emit 
photons in forward direction $\f{\kappa}$. 
Only in this case, all the phases $e^{i \fk{\kappa} \cdot \fk{r}_\ell}$ 
add up coherently (we assume random locations $\f{r}_\ell$),
see Fig.~\ref{fig:exciton_and_dicke}.
We will now investigate spontaneous and stimulated emission from an 
ensemble of $S$ coherently excited nuclei in more detail.
\subsection{Spontaneous emission}
We start with the case of collective spontaneous emission 
(a.k.a.\ Dicke super-radiance \cite{Scully:2006fk,Scully:2007fk,Sete:2010fu}) 
from a coherent state $\ket{s}$. 
First of all, as the $S$ nuclei are not enclosed by a resonator 
or a cavity in our set-up, we have to consider all $\f{k}$-modes. 
Thus, the Hamiltonian~(\ref{eq:hinteract}) changes into
\bea
\label{eq:hmanymodes}
\hat{V}_{\rm sp} \left( \tau \right) 
= 
\int d^3k\; g_{\fk{k}} \hat{a}_{\fk{k}} \, 
e^{-i \left( \omega_{\fk{k}} - \omega \right) \tau} \hat{\Sigma}^+ 
\left( \f{k} \right) + {\rm H.c.}
\,,
\ea
where $\hat{a}_{\fk{k}}$ is the photonic annihilation operator for the
mode $\f{k}$ with frequency $\omega_{\fk{k}}$ and $g_{\fk{k}}$ the 
associated coupling strength.
Note that we neglect polarization effects, i.e., we assume that 
the polarization vectors are directed along the same axis as the dipole 
moments of the absorbing nuclei.
Furthermore, $\hat{\Sigma}^+\left(\f{k}\right)$ denotes the 
quasispin-$S$-operators with the wavenumber $\f{k}$ instead of $\f{\kappa}$. 
However, when $\f{k}$ is not close to $\f{\kappa}$, the phase factors 
of $\hat{\Sigma}^\pm \left( \f{k} \right)$ and $\ket{s}$ do not match, 
and the transition is not coherent, i.e., not enhanced by a factor 
$S$ according to Eq.~(\ref{eq:matrix_elements}), 
and can thus be neglected. 
Note that this is the reason why collectively emitted photons are 
directed along (almost) the same axis as previously absorbed 
photons \cite{Scully:2006fk,Scully:2007fk,Sete:2010fu}, 
see also Fig.~\ref{fig:exciton_and_dicke}.
For simplicity, the quasispin-$S$-operators are therefore approximated 
by introducing a cut-off function $g \left( \f{\kappa} - \f{k} \right)$ 
that is only non-zero for small deviations $\f{\kappa} - \f{k}$ 
of the supported direction $\f{\kappa}$, i.e., 
$\hat{\Sigma}^\pm \left( \f{k} \right) \approx 
g \left( \f{\kappa} - \f{k} \right)\,\hat{\Sigma}^\pm$.
For large $S$, we may approximate the quasispin-$S$-operators 
classically, i.e., 
$\hat{\Sigma}^- \approx \Sigma^- = \sqrt{\left( S - s +1 \right) s}$, 
and thus the effect of the Hamiltonian Eq.~(\ref{eq:hmanymodes}) acting on 
the vacuum state can be expressed by a coherent state
\bea
\hat{U}_{\rm sp} ( t ) \ket{0} 
\approx 
\exp\left( 
\int d^3k \, \alpha_{\fk{k}} \hat{a}_{\fk{k}}^\dagger - {\rm H.c.}
\right) 
\ket{0}
\,,
\ea
with the amplitudes 
\bea
i\alpha_{\fk{k}} 
= 
g_{\fk{k}}^* g ( \f{\kappa} - \f{k} ) 
\sqrt{\left( S - s +1 \right) s} 
\int_{0}^t
d\tau \, 
e^{i \left( \omega_{\fk{k}} - \omega \right) \tau}
\,.
\ea
The number of emitted photons per mode is given by $|\alpha_{\fk{k}}|^2$
and the total photon number grows linearly with $t$
\bea
\mathcal{N}_\gamma 
= 
\int d^3k \, |\alpha_{\fk{k}}|^2 
\approx 
2 \pi^2 \left( S - s +1 \right) s |g_{\fk{\kappa}}|^2 \frac{t}{L_\bot^2}
\,,
\ea
where $L_\bot^2$ denotes the transversal cross-section area of the ensemble,
which determines the  transverse area in $\f{k}$-space where 
$g ( \f{\kappa} - \f{k} )$ is non-zero.
In addition to this spatial resonance condition, the temporal resonance 
was incorporated via approximating the squared time-integral by $t^2$ for 
$| \omega_k - \omega | < 1/t$ and zero otherwise. 

Strictly speaking, this relation is only valid for a fixed number of 
excitations $s$, i.e., the time-dependence of $s(t)$ due to the emission 
of photons (energy conservation) is neglected. 
Assuming that this time-dependence $s(t)$ is slow compared to $\omega$ 
(i.e., that the coupling strength is small enough), we may take it into 
account approximately via defining the instantaneous spontaneous emission 
rate 
\bea
\Gamma_{\rm sp}(t) 
= \frac{d\mathcal{N}_\gamma}{dt} 
= \gamma \left[ S - s(t) +1 \right] s(t)
\,,
\ea
with the abbreviation $\gamma = 2 \pi^2 |g_{\fk{\kappa}}|^2 / L_\bot^2$. 
The change of $s(t)$ in the time interval $dt$ is then governed by 
$\Gamma_{\rm sp}(t)$
\bea
\label{eq:dgl}
\frac{ds(t)}{dt} 
= 
- \Gamma_{\rm sp}(t) 
= 
-\gamma \left[ S - s(t) + 1 \right] s(t)
\,.
\ea
For the initial condition $s(0) = S/2$ (see below), 
the solution for $S \gg 1$ is given by 
\bea
s(t) = \frac{S}{1 + e^{\gamma S  t}}
\,.
\ea
This yields the intensity due to spontaneous emission 
\bea
\label{eq:ispon}
I_{\rm sp}(t) 
&=& 
-\frac{dE}{dt} \frac{1}{L_\bot^2} 
= 
-\frac{ds(t)}{dt} \frac{\omega}{L_\bot^2} 
\nonumber\\
&=& 
\frac{1}{4} \gamma S^2 {\rm sech}^2 \left( \gamma S t /2 \right) 
\frac{\omega}{L_\bot^2}
\,,
\ea
where $L_\bot^2$ is the cross-section area of the emitted beam, 
see also \cite{Rehler:1971fk}. 
The time-dependence in the ${\rm sech}$-function can be used to define 
an effective time constant via 
\bea
\label{eq:tauspon}
\tau_{\rm sp} = \frac{4}{\gamma S} 
\,, 
\ea
after which $I_{\rm sp}(t)$ has dropped to 7\% of its initial value. 
Let us now briefly compare this time-scale $\tau_{\rm sp}$ for the coherent 
spontaneous emission process with the time-scale in the incoherent case. 
For incoherent emission from $s$ excited nuclei, we can regard each 
nucleus independently. 
According to standard Weisskopf-Wigner theory \cite{Scully:1997fk},  
the life-time of an excited nucleus is given by
\bea
\label{eq:tausingle}
\tau_{\rm single} 
= 
\frac{1}{\Gamma_{\rm single}} 
= 
\frac{1}{8 \pi^2} \frac{1}{|g_{\fk{\kappa}}|^2 \omega^2}
\,.
\ea
Comparing the two time-scales Eq.~(\ref{eq:tauspon}) and (\ref{eq:tausingle})
\bea
\frac{\tau_{\rm sp}}{\tau_{\rm single}}
=
64\pi^2 \frac{L_\bot^2}{\lambda^2} \frac{1}{S}
\,,
\ea
we find that for large $S$, the coherent spontaneous emission process is 
much faster than the incoherent process 
(see also \cite{Junker:2012fk}). 
Taking for example $S = 10^{10}$ $^{57}$Fe-nuclei with resonance at 
$\Delta E_\gamma = 14.4\;{\rm keV}$, lifetime 
$\tau_{\rm single} = 141\;{\rm ns}$ and $L_\bot = 0.1\;\mu{\rm m}$, 
the quotient evaluates to $\tau_{\rm sp}/\tau_{\rm single} \approx 0.09$, 
i.e. the coherent emission runs over ten times faster than the incoherent 
emission.
Note that there are also competing processes, such as decay via 
electron conversion -- but they are incoherent and thus can be suppressed 
for large $S$, i.e., small $\tau_{\rm sp}$.

\subsection{Stimulated emission}
In order to study stimulated emission from a coherently excited $S$-nuclei 
ensemble, we regard the incoming field 
$A_{\rm in}(t) = \sqrt{I_{\rm in}(t)}/\omega$ classically. 
That is, we use the Hamiltonian Eq.~(\ref{eq:hinteract}), but replace 
$g \hat{a}$ by $\tilde{g} A_{\rm in}(t)$. 
For simplicity, we assume the transition matrix element $\tilde{g}$ 
of the nucleus to be real
\bea
\label{eq:hstim}
\hat{V}_{\rm st} 
= 
\tilde{g} A_{\rm in}(t) \left( \hat{\Sigma}^+ + \hat{\Sigma}^- \right) 
= 
2 \tilde{g} A_{\rm in}(t) \hat{\Sigma}^x
\,.
\ea
Applying Heisenberg picture and employing the properties of the 
$SU(2)$-algebra yields
\bea
\label{eq:sigmaz}
\hat{U}_{\rm st}^\dagger (t) \hat{\Sigma}^z \hat{U}_{\rm st} (t) 
&=& 
\cos \left( 2 \tilde{g} \int_0^t d\tau\, A_{\rm in}(\tau)\right) 
\hat{\Sigma}^z
\nonumber\\
&+& \sin \left( 2 \tilde{g} \int_0^t d\tau\, A_{\rm in}(\tau)\right) 
\hat{\Sigma}^y
\,.
\ea
As envisaged for laser application (see below), we choose $s(0) = S$ here, 
that is all $S$ nuclei are in the coherently excited state. 
The time-dependent number of excitations is given by 
$\bra{S}\hat{U}_{\rm st}^\dagger(t)\hat{\Sigma}^z\hat{U}_{\rm st}+S/2\ket{S}$ 
and thus the energy stored in the $S$ nuclei at time $t$ is
\bea
E(t) 
= 
\frac{S \omega}{2} \left[ \cos 
\left( 2 \tilde{g} \int_0^t d\tau\, A_{\rm in}(\tau)\right) 
+ 1\right]
\,.
\ea
This yields the emitted intensity $I_{\rm st}(t)$ stimulated by the incoming 
intensity $I_{\rm in}(t)$ 
\bea
\label{eq:istim}
I_{\rm st}(t) 
= 
\frac{\tilde{g} S}{L_\bot^2} 
\sin \left( 
\frac{2 \tilde{g}}{\omega} \int_0^t d\tau\, \sqrt{I_{\rm in}(\tau)}
\right) 
\sqrt{I_{\rm in}(t)}
\,,
\ea
where we have assumed that both beams have the same cross-section area 
$L_\bot^2$. 

We define the time-scale of the stimulated emission as the time 
$\tau_{\rm st}$, after which all the energy initially stored in the $S$ 
nuclei has been emitted, i.e., 
\bea
\label{eq:taustim}
\int_0^{\tau_{\rm st}} d\tau\,\sqrt{I_{\rm in}(\tau)} 
= 
\frac{\pi \omega}{2 \tilde{g}}
\,.
\ea
Now let us imagine that we have two separate ensembles (e.g., foils) 
of coherently excited nuclei, such that the first foil spontaneously
emits the intensity $I_{\rm in}(t)$ as in Eq.~(\ref{eq:ispon}) which 
causes stimulated emission according to Eq.~(\ref{eq:istim}) 
in the second foil. 
In this case, we can insert $ I_{\rm in}(t) = I_{\rm sp}(t)$, 
and Eq.~(\ref{eq:taustim}) can be solved for $\tau_{\rm st}$
\bea
\tau_{\rm st}  
= 
\frac{4}{\gamma S}\,{\rm ArTanh}
\left[ {\rm Tan} \left( \frac{1}{8} 
\sqrt{\frac{\pi}{2}} \right) \right] 
\approx 0.16\times\tau_{\rm sp}
\,.
\ea
Since both foils contain the same nuclei (with the same coupling strengths), 
the time-scale for the stimulated emission of the second foil, 
$\tau_{\rm st}$, is completely determined by the time-scale of the 
spontaneous emission process of the first foil, $\tau_{\rm sp}$.

\section{Pumping}
After having discussed coherent spontaneous emission as well as coherent 
stimulated emission, let us investigate the pumping process for a single 
foil of $S$ nuclei, which are initially in the state $s=0$, i.e., 
$\bra{0} \hat{\Sigma}^z \ket{0} = -S/2$. 

Note that it is very easy to over- or under-estimate the efficiency of the 
pumping process by using too simplified pictures.
On the one hand, one might expect that the number of excitons in the foil 
grows linearly with the number of photons incident and thus linearly with 
the interaction time $t$. 
However, this is only true for pumping with incoherent light 
(for further details, see the appendix), but not for coherent pumping, 
which is the case considered here. 
On the other hand, since the transition matrix elements in 
Eq.~(\ref{eq:matrix_elements}) scale with $\sqrt{s}$ and thus the 
effective line-width increases with $s$, one might expect a behavior 
like $\dot s\propto s$, which would imply an exponential growth 
$s(t)\propto e^{\kappa t}$, at least for small $s\ll S$. 
This picture is also wrong, since -- in view of the unitarity of the 
time-evolution -- not just the absorption rate but also the emission 
rate increase with $s$.  
Thus, the correct answer is that $s(t)$ grows quadratically 
$s(t)\propto t^2$ for small $s$, i.e., somewhere in between 
linear and exponential. 

To show this, let us consider pumping with one coherent pump-pulse 
$A_{\rm pump}(t)$ for the whole interaction time. 
We can employ the Hamiltonian Eq.~(\ref{eq:hstim}) again with the sole 
difference that the incoming field $A_{\rm in}(t)$ is now given by the 
pump-field $A_{\rm in}(t) = A_{\rm pump}(t)$. 
Thus, Eq.~(\ref{eq:sigmaz}) again holds, and the exciton number is given by 
\bea
s(t) 
&=&
\frac{S}{2} 
\left[ 
1 - \cos \left( 2 \tilde{g} \int_0^t d\tau\, A_{\rm pump}(\tau)\right) 
\right]
\nonumber\\
&=&
S \tilde{g}^2 \left( \int_0^t d\tau\, A_{\rm pump}(\tau) \right)^2 
+ \ord \left( \tilde{g}^4 t^4 \right)
\,,
\ea
i.e., the exciton number grows quadratically for small $t$ 
(for an alternative approach, see the appendix). 
We moreover find that a full cycle 
(i.e., a sign flip of $\hat{\Sigma}^z\to-\hat{\Sigma}^z$) 
occurs after the pump time $\tau_{\rm pump}$ where 
\bea
\label{eq:pump_constraint}
\int_0^{\tau_{\rm pump}} d\tau\, A_{\rm pump}(\tau) = \frac{\pi}{2 \tilde{g}}
\,.
\ea
The simplest example would be a constant pump pulse 
$A_{\rm pump}=A_0$ with $\tau_{\rm pump} = \pi / ( 2 \tilde{g} A_0 )$.
In order to see if such a pump-field is feasible in general, we calculate 
a rough estimate for the required intensity of the pump-field. 
For simplicity, we assume that the intensity is constant over the pulse, 
i.e., $A_{\rm pump} (t) = \sqrt{I_{\rm pump}}/\omega$. 
Then we find 
$I_{\rm pump}  = \pi^2 \omega^2 / (4 \tilde{g}^2 \tau_{\rm pump}^2)$. 
Now $\tilde{g}$ can be expressed in terms of the (single-nucleus) decay 
rate $\Gamma_{\rm single}$ and the frequency $\omega$ of the considered 
nuclear excitation (see the appendix). 
Moreover, if we replace the coherent pulse-length 
$\tau_{\rm pump} = {\mathfrak N} \lambda = 2 \pi {\mathfrak N} / \omega$ 
by the number ${\mathfrak N}$ of (coherent) wave-cycles, we find 
\bea
\label{eq:ipump}
I_{\rm pump}  
= 
\frac{1}{32 \pi {\mathfrak N}^2}\,\frac{\omega^5}{\Gamma_{\rm single}}
\,.
\ea
Thus, nuclear resonances with low energies $\omega$ but high decay-rates 
$\Gamma_{\rm single}$ require low pump intensities.
Concrete examples will be discussed at the end of this article.

\section{Laser}
\begin{figure}[h]
\begin{center}
\subfigure[]{\includegraphics[width=0.4\columnwidth]{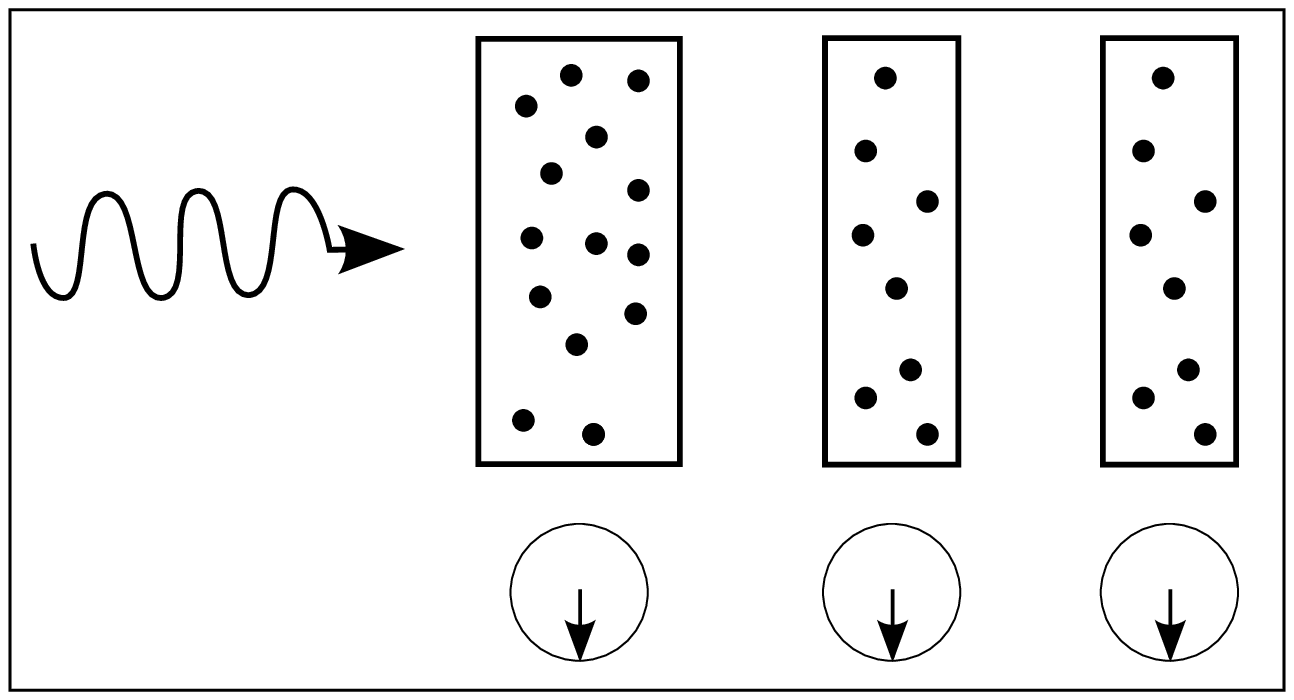}}
\hspace{.5cm}
\subfigure[]{\includegraphics[width=0.4\columnwidth]{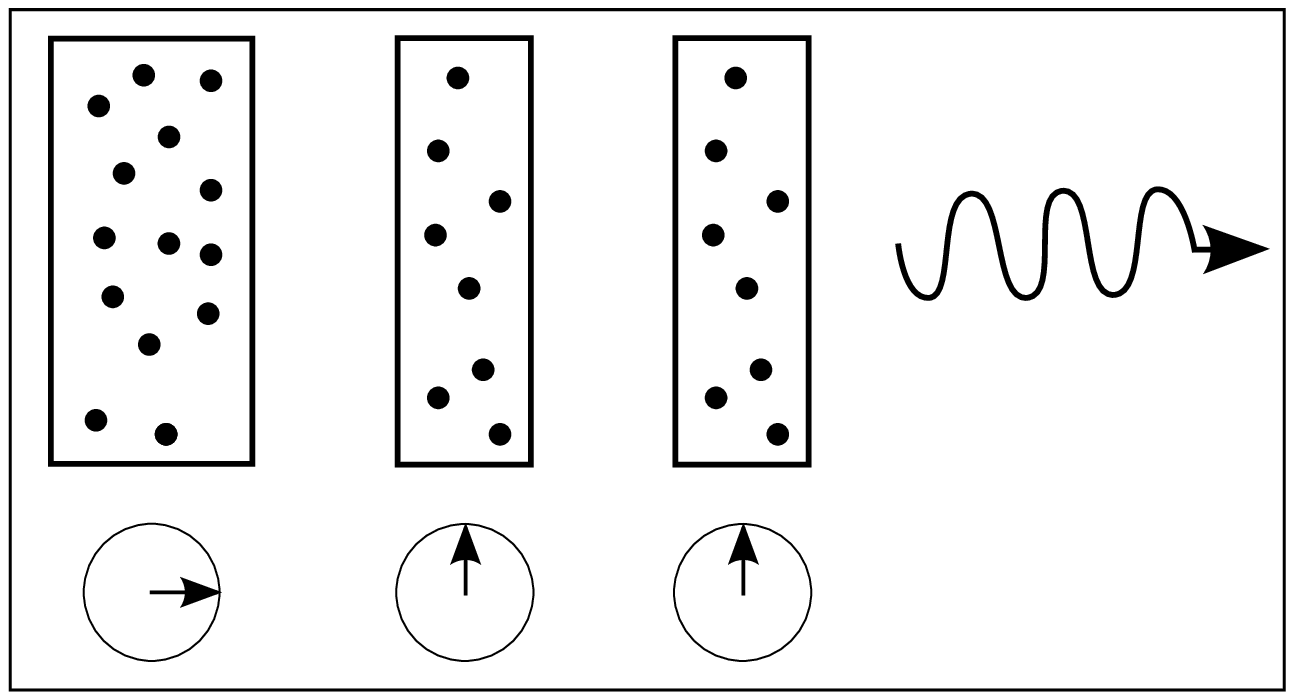}}
\subfigure[]{\includegraphics[width=0.4\columnwidth]{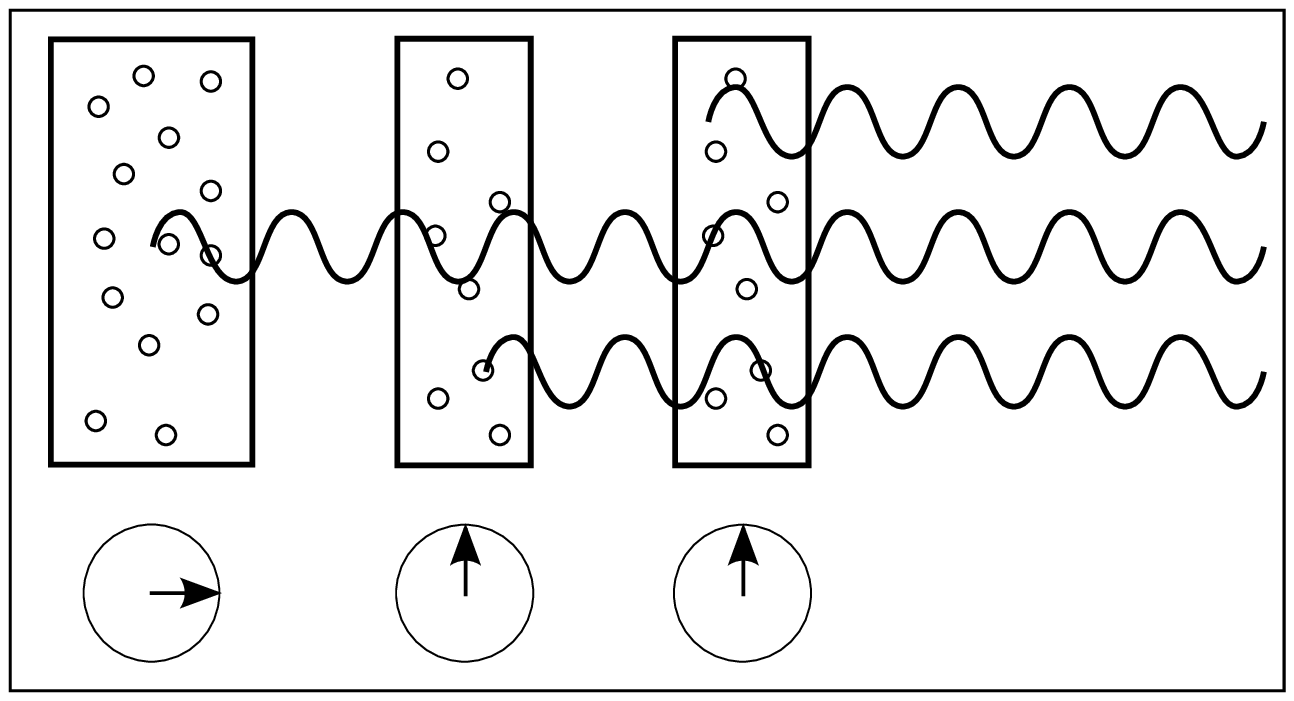}}
\caption{
Sketch of the operation sequence of the proposed nuclear exciton laser.
Initially, all foils (here $N=3$) are in the ground state 
$\Sigma^z_n = -S_n/2$ (a).
The pump-pulse then rotates the quasispin of the first foil to 
$\Sigma^z_1 = 0$ 
and the quasispin of all subsequent foils to $\Sigma^z_n = +S_n/2$ (b). 
Then, the ``half-filled'' first foil spontaneously emits a pulse 
$I_{\rm sp}(t)$, 
which stimulates emission at foils $2$ and $3$, leading to an enhanced 
overall intensity $I_{\rm total}^{(3)}(t)$ (c).
}
\label{fig:sequence}
\end{center}
\end{figure}

Now we have gathered all the tools required to understand the set-up 
of the proposed nuclear exciton laser.
The envisaged set-up consists of a series of $N \gg 1$ foils 
$n = 1,2,...,N$, each foil containing $S_n$ nuclei 
(two-level systems) with a nuclear resonance at frequency $\omega$. 
At the beginning, we assume that all $n = 1,2,...,N$ foils are in the 
ground state, corresponding to a quasi-spin $\Sigma^z_n = -S_n/2$, 
see Fig.~\ref{fig:sequence}(a).

To prepare the emission of a laser pulse, the foils need to be pumped 
to suitable coherent states. 
Let us distinguish between the first foil and all later foils $n = 2,...,N$. 
While the latter all should be pumped to the maximum of $\Sigma^z_n = S_n/2$, 
i.e., $s_n = S_n$, the first foil should only be pumped such that half of the 
nuclei are in the excited state, i.e., $\Sigma^z_1 = 0$ and $s_1 = S_1/2$. 
For simplicity, we envisage the whole pumping process to be achieved by only 
one coherent pump-pulse, which goes through all the foils one after another
and is only weakly changed by absorption. 
The pump-pulse should satisfy Eq.~(\ref{eq:pump_constraint}) in order to 
rotate the quasispin $\Sigma^z_n$ of each foil from $\Sigma^z_n = -S_n/2$ 
to $\Sigma^z_n = +S_n/2$. 
Additional measures need to be taken to ensure that the first foil is only 
pumped to $s_1 = S_1/2$. 
One option could be to have a different kind of nuclei in the first foil,
which have the same resonance frequency as those in the other foils, while 
the coupling strengths differ by a factor of two (approximately).
Another option could be to switch the first foil
(mechanically or magnetically 
\cite{Shvydko:1996fk,Rohlsberger:2000fk,Coussement:2002zr,
Palffy:2009kx,Adams:2011ys}) 
during the pumping process.  

When the set-up is prepared as shown in Fig.~\ref{fig:sequence}(b), 
the emission process automatically starts, as the first foil immediately 
begins with the spontaneous emission discussed above, Eq.~(\ref{eq:ispon}). 
The idea is that, due to the ``half-filled'' coherent state $\Sigma^z_1 = 0$, 
the emission process of the first foil happens much faster than the 
spontaneous emission of the subsequent foils. 
Taking, e.g., the second foil ($s_2 = S_2$), the time-scale for the emission 
of a single photon would be $1 / \Gamma_{\rm sp} = 1 / (\gamma S_2)$. 
For the first foil, the time-scale for the whole emission process 
(of nearly all photons, not only one) is given by 
$\tau_{\rm sp} = 4 / (\gamma S_1)$. 
So by choosing $S_1 \gg S_2$, e.g., by making the first foil ten times 
thicker than the subsequent foils, it is assured that the second foil 
is still in the state $\Sigma^z_2 = S_2/2$, when the intensity emitted 
from the first foil is incident.

Stimulated emission then occurs at the second foil according to 
Eq.~(\ref{eq:istim}) and the second foil has emitted all its energy 
after $\tau _{\rm st} \approx 0.16\times\tau_{\rm sp}$, i.e., 
before the stimulating pulse coming from the first foil declines. 
After the second foil, the overall intensity thus adds up to 
$I_{\rm total}^{(2)}(t) = I_{\rm sp}(t) + I_{\rm st}^{(2)}(t)$.
This overall intensity then causes stimulated emission at the third foil, 
resulting in an even bigger intensity 
$I_{\rm total}^{(3)}(t) = I_{\rm total}^{(2)}(t) + I_{\rm st}^{(3)}(t)$, etc. 
In this way, the intensity of the light pulse grows stepwise with each 
passed foil.

Numerical analysis has been done for the case of $N = 50$ foils. 
Iteratively, $I_{\rm st}^{(n)} (t)$ was calculated from 
$I_{\rm total}^{(n-1)} (t)$, where 
$I_{\rm total}^{(n)} (t) = I_{\rm total}^{(n-1)} (t) + I_{\rm st}^{(n)} (t)$, 
starting with $I_{\rm total}^{(1)} (t) = I_{\rm sp}(t)$. 
It was assumed that the first foil consists of $S_1 = 10^{10}$ 
$^{57}$Fe-nuclei 
(with $\Delta E_\gamma = 14.4\;{\rm keV}$ and $\tau = 141\;{\rm ns}$)  
while all other foils are ten times thinner, i.e., $S_n = 10^{9}$.
Transversal dimensions of the foils and the laser beam are chosen as  
$L_\bot^2 = (0.1\;\mu{\rm m})^2$.

\begin{figure}[h]
\begin{center}
\psfrag{Itotaltau}
{{$\overline{I_{\rm total}^{(n)}}/\overline{I_{\rm total}^{(2)}}$}}
\psfrag{n}{{$n$}}
\includegraphics[width=0.9\columnwidth]{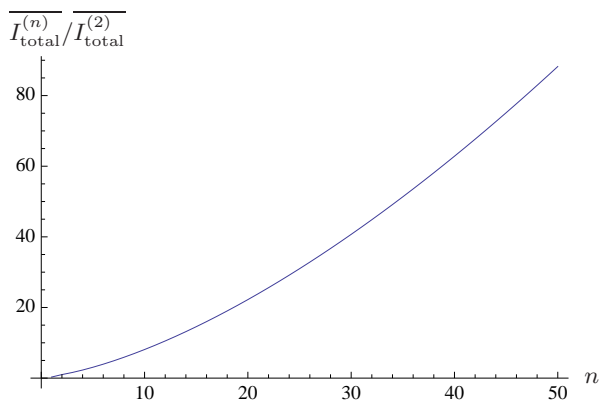}
\caption{Average intensity $\overline{I_{\rm total}^{(n)}}$ over number of 
foil $n$ from numerical analysis. In this example, 
$\overline{I_{\rm total}^{(n)}}$ roughly increases as a power law 
$\propto n^{3/2}$.}
\label{fig:powerlaw}
\end{center}
\end{figure}

Note that the useful part of the laser pulse $I_{\rm total}^{(n)}(t)$ 
is determined by the time after which the last foil has emitted all its 
excitations, $\tau^{(n)}_{\rm st}$, because afterwards re-absorption 
takes place. 
This time $\tau^{(n)}_{\rm st}$ becomes shorter with rising $n$, 
as the intensity which causes the stimulated emission grows with $n$.
As a result, the average intensity of the useful part of the laser pulse, 
\bea
\overline{I_{\rm total}^{(n)}} 
= 
\frac{1}{\tau^{(n)}_{\rm st}}
\int_0^{\tau^{(n)}_{\rm st}} d\tau\, 
I_{\rm total}^{(n)} (\tau)
\,, 
\ea
increases with a power law. In the concrete example given above, 
$\overline{I_{\rm total}^{(n)}}$ grows roughly $\propto n^{3/2}$, 
see Fig.~\ref{fig:powerlaw}.

\bigskip

\section{Conclusions}
In summary, we described a proposal for a laser in the $\ord({\rm keV})$
regime which is based on stimulated emission and works with nuclear 
excitons. 
The pumping could be achieved with a free-electron laser, for example.
Note that the pump pulse $A_{\rm pump}$ and the generated laser pulse 
$A_{\rm laser}$ both correspond to a 180$^\circ$-rotation of the last 
foil according to Eq.~(\ref{eq:sigmaz}) and thus are related via  
\bea
\int_0^{\tau_{\rm pump}} d\tau\, A_{\rm pump}(\tau) 
=
\int_0^{\tau^{(N)}_{\rm st}} d\tau \, A_{\rm laser}(\tau)
=
\frac{\pi}{2\tilde g}
\,. 
\ea
However, the intensity of the pump pulse $\propto|A_{\rm pump}^2|$ is much 
larger than that of the laser pulse $\propto|A_{\rm laser}^2|$.
On the other hand, the duration ${\tau^{(N)}_{\rm st}}$ of the laser pulse
is much larger and thus its frequency accuracy is much higher
(see also \cite{Kim:2008qf} for a different approach).
This could be important for spectroscopy etc. 

Let us discuss some example data for the required intensity of the 
pump-pulse. 
First, we consider $^{57}$Fe-nuclei with a resonance at 
$\Delta E_\gamma = 14.4\;{\rm keV}$ with a mean lifetime of 
$\tau = 141\;{\rm ns}$. 
If we assume that the pump pulse consists of ${\mathfrak N} = 10^6$ 
coherent wave-trains, we would need a pump intensity of 
$I_{\rm pump} \approx 8.3 \cdot 10^{20} \;{\rm W}/{\rm cm^2}$
according to Eq.~(\ref{eq:ipump}). 
(Comparable or even higher intensities have already been 
considered in e.g.\ \cite{Burvenich:2006bh,Palffy:2008uq,Liao:2011cr}.)
This is probably beyond the capabilities of present 
free-electron lasers, see, e.g.,  \cite{felbasics}.  
However, future light sources such as seeded free-electron lasers 
should achieve improved coherence times and higher intensities 
(especially after focussing with X-ray lenses). 

On the other hand, when considering other nuclear resonances beyond 
the well-known $^{57}$Fe-example, we find that the requirements are 
somewhat easier to fulfill.
For example, considering the $^{201}$Hg resonance at 
$\Delta E_\gamma = 1.6\;{\rm keV}$ with $\tau = 81\;{\rm ns}$
and again assuming ${\mathfrak N} = 10^6$ coherent wave-trains,
we would ``only'' need a pump intensity of 
$I_{\rm pump} \approx 8.0 \cdot 10^{15}\;{\rm W}/{\rm cm^2}$ 
according to Eq.~(\ref{eq:ipump}).

Unfortunately, this intensity is probably still too large:
After inserting typical values for the absorption cross section of 
$1.6\;{\rm keV}$-photons in metals (or other solid materials), we find 
that the pump beam deposits enough energy in the foil to evaporate it.  
Even though the thermalization dynamics following the illumination 
with such a $8.0 \cdot 10^{15}\;{\rm W}/{\rm cm^2}$-beam of 
$1.6\;{\rm keV}$-photons is not well studied yet, one would expect that 
the foil starts to disintegrate after a few pico-seconds \cite{Siwick:2003fk}
and hence does not survive long enough for our purposes.

In summary, the major difficulty of our set-up is that it requires 
extremely large pump intensities. 
As we may infer from Eq.~(\ref{eq:ipump}), the pump intensity scales with 
the fifth power of the transition energy $\omega$. 
Thus, our scheme should be much easier to realize at lower energies. 
As one possible example, let us envisage a UV-laser.
In this case, the nuclear transitions could be replaced by 
suitable electronic transitions in atoms or molecules.
The pumping could be achieved either directly via a free-electron 
laser in the low-energy regime or indirectly via a two-photon transition 
generated by two optical lasers, for example. 

\begin{figure}[h]
\begin{center}
\psfrag{e1}{$E_2$}
\psfrag{e2}{$E_3$}
\psfrag{e3}{$E_4$}
\psfrag{1s}{2s}
\psfrag{2s}{3d}
\psfrag{3p}{4p}
\psfrag{delta}{$\Delta$}
\psfrag{omega_1}{$\omega_1$}
\psfrag{omega_2}{$\omega_2$}
\psfrag{omega_uv}{$\omega_{\rm uv}$}
\includegraphics[width=0.5\columnwidth]{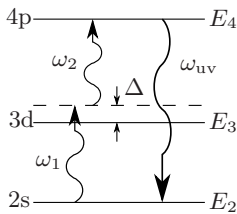}
\caption{Sketch (not to scale) of the level scheme. 
The pumping process from 2s to 4p is induced by a detuned two-photon 
transition, i.e., both photons together are in resonance 
$E_4-E_2=\omega_1+\omega_2$ while one-photon absorption is 
suppressed by the detuning $\Delta$ where $E_3-E_2=\omega_1-\Delta$. 
The laser operates via the one-photon transition from 4p back to 2s and 
emits photons of the energy $\omega_{\rm uv}=\omega_1+\omega_2$.} 
\label{fig:three_level}
\end{center}
\end{figure}

Let us discuss the latter case using a three-level system as 
depicted in Fig.~\ref{fig:three_level}. 
Assuming two pump-lasers with optical frequencies $\omega_1$ and 
$\omega_2$, the laser could operate in the ultra-violet regime  
$\omega_{\rm uv}$.
In this case, the expression $\tilde g A_{\rm pump}$ for one-photon 
pumping in Eq.~(\ref{eq:pump_constraint}) should be replaced by 
$\tilde g_{23}A^{\rm pump}_1\tilde g_{34}A^{\rm pump}_2/\Delta$. 
Note that the coupling constant $\tilde g_{23}$ of the 
``dipole-forbidden'' 2s-3d transition 
is typically much smaller than $\tilde g_{34}$. 
Assuming typical values, such as a dipole coupling length of three Bohr 
radii,
we would need pump-laser intensities of about 
$I_{\rm pump} = \ord(10^{10}\;{\rm W}/{\rm cm}^2)$ 
over a length of ${\mathfrak N} = 10^4$ coherent wave-trains with a 
detuning of $\Delta = \ord(10^{13}\;{\rm Hz})$ in order to
prevent unwanted excitations of the middle 3d level.
The condition for dominant coherent emission, 
$\tau_{\rm sp}/\tau_{\rm single} \ll 1$, 
can be fulfilled for $S_1/L_\bot^2 = \ord(10^{5}\;\mu{\rm m}^{-2})$, 
which is quite reasonable.
In this scenario, the duration of the laser pulse ${\tau^{(N)}_{\rm st}}$ 
is comparable to the length of the pump pulse 
$\tau_{\rm pump} \approx {\tau^{(N)}_{\rm st}} = \ord(10\;{\rm ps})$
and its intensity is well above $\ord(10^5\;{\rm W}/{\rm cm}^2)$,
depending on the number of foils.  

Again, the main idea would be that the coherent emission is strongly 
enhanced for $S\gg1$ in comparison to competing non-coherent decay 
channels.
To this end, the two pump lasers must be parallel to ensure the 
spatial phase matching.

\section*{Acknowledgements}
R.S.\ acknowledges fruitful discussions, e.g., with Bernhard Adams, 
at the Winter Colloquia on the Physics of Quantum Electronics (PQE);
and with Uwe Bovensiepen. 
N.t.B.\ would like to thank Sumanta Das for valuable discussions at the
DPG Spring Meeting in Stuttgart.
This work was supported by the DFG (SFB-TR12). 

\section*{Appendix}
\subsection{Coherent versus incoherent pumping}
Let us review the pumping process by applying the 
Holstein-Primakoff \cite{Holstein:1940vn} transformation
\bea
\hat{\Sigma}^+ 
= 
\hat{b}^\dagger \sqrt{S - \hat{b}^\dagger \hat{b}} 
= 
(\hat{\Sigma}^-)^\dagger
\,,\quad 
\hat{\Sigma}^z = \hat{b}^\dagger \hat{b} - S/2
\,,
\ea
to the Hamiltonian Eq.~(\ref{eq:hstim}) and considering the limit $S\gg s$, 
i.e., the beginning of the pumping process
\bea
\hat{V}_{\rm st} 
\approx 
\tilde{g} A_{\rm pump}(t) 
\left( \sqrt{S} \hat{b}^\dagger + {\rm H.c.} \right)
\,.
\ea
We first analyze the case of coherent pumping, that is pumping with a 
coherent pulse $A_{\rm pump}(t)$, i.e., the same time-evolution operator 
for the whole pumping process
\bea
\hat U_{\rm st}(t) = \exp\left(
\beta(t)\hat b^\dagger-{\rm H.c.}
\right)
\,.
\ea
A time-dependent coherent state of excitons is created 
\bea
\beta(t)=-i \tilde{g} \sqrt{S} \int_0^t d\tau A_{\rm pump}(\tau)
\,,
\ea
whose exciton number grows quadratically with time $t$
\bea
n(t)=|\beta(t)|^2
=\ord\left( \tilde{g}^2 S A_{\rm pump}^2 t^2 \right)
\,.
\ea

An incoherent pump-pulse, in contrast, can be approximated as a 
succession of many uncorrelated coherent pulses $A_{\rm pump}^{(i)}(t)$ 
incident on the target. 
The time-evolution operator then is a product of many coherent 
displacement operators 
\bea
\hat U_{\rm eff}
\approx
\prod_i
\hat{U}_{\rm st}^{(i)}
=
\exp\left(\sum_i\beta_i \hat b^\dagger-{\rm H.c.}\right)
\,.
\ea
For uncorrelated pulses, the $\beta_i$ have random phases, 
such that the sum corresponds to a random walk
\bea
\beta_{\rm eff}(j)
=\sum_{i=1}^j\beta_i\propto\sqrt{j}\propto\sqrt{t}
\,,
\ea
such that the exciton number $n = |\beta_{\rm eff}|^2$ 
grows merely linearly with time in this case.

\subsection{Expressing $\tilde{g}$ in terms of $\Gamma_{\rm single}$ and $\omega$}
The coupling constant $|g_{\fk{\kappa}}|$ can be expressed via the decay 
rate $\Gamma_{\rm single}$ and frequency $\omega$
\bea
\Gamma_{\rm single} 
= 
2 \pi \int d^3k\, |g_{\fk{k}}|^2 \delta 
\left( \omega_k - \omega \right) 
\approx 
8 \pi^2 |g_{\fk{\kappa}}|^2 \omega^2
\,.
\ea
The dimensionless coupling constant $\tilde{g}$ can be obtained via 
$\tilde{g} = |g_{\fk{\kappa}}| \sqrt{2 (2 \pi)^3 \omega}$ since our 
Hamiltonian contains the classical field $A_{\rm pump}(t)$.
As a result we arrive at
\bea
\tilde{g} = \sqrt{\frac{2 \pi \Gamma_{\rm single}}{\omega}}
\,.
\ea

\bibliographystyle{apsrev}
\bibliography{e11}

\begin{thebibliography}{34}
\expandafter\ifx\csname natexlab\endcsname\relax\def\natexlab#1{#1}\fi
\expandafter\ifx\csname bibnamefont\endcsname\relax
  \def\bibnamefont#1{#1}\fi
\expandafter\ifx\csname bibfnamefont\endcsname\relax
  \def\bibfnamefont#1{#1}\fi
\expandafter\ifx\csname citenamefont\endcsname\relax
  \def\citenamefont#1{#1}\fi
\expandafter\ifx\csname url\endcsname\relax
  \def\url#1{\texttt{#1}}\fi
\expandafter\ifx\csname urlprefix\endcsname\relax\def\urlprefix{URL }\fi
\providecommand{\bibinfo}[2]{#2}
\providecommand{\eprint}[2][]{\url{#2}}

\bibitem[{\citenamefont{Baldwin and Solem}(1997)}]{Baldwin:1997ve}
\bibinfo{author}{\bibfnamefont{G.~C.} \bibnamefont{Baldwin}} \bibnamefont{and}
  \bibinfo{author}{\bibfnamefont{J.~C.} \bibnamefont{Solem}},
  \bibinfo{journal}{Rev. Mod. Phys.} \textbf{\bibinfo{volume}{69}},
  \bibinfo{pages}{1085} (\bibinfo{year}{1997}).

\bibitem[{\citenamefont{Tkalya}(2011)}]{Tkalya:2011dq}
\bibinfo{author}{\bibfnamefont{E.~V.} \bibnamefont{Tkalya}},
  \bibinfo{journal}{Phys. Rev. Lett.} \textbf{\bibinfo{volume}{106}},
  \bibinfo{pages}{162501} (\bibinfo{year}{2011}).

\bibitem[{\citenamefont{M{\"o}ssbauer}(1958{\natexlab{a}})}]{Mossbauer:1958fk}
\bibinfo{author}{\bibfnamefont{R.~L.} \bibnamefont{M{\"o}ssbauer}},
  \bibinfo{journal}{Z. Physik} \textbf{\bibinfo{volume}{151}},
  \bibinfo{pages}{124} (\bibinfo{year}{1958}{\natexlab{a}}).

\bibitem[{\citenamefont{M{\"o}ssbauer}(1958{\natexlab{b}})}]{Mossbauer:1958kx}
\bibinfo{author}{\bibfnamefont{R.~L.} \bibnamefont{M{\"o}ssbauer}},
  \bibinfo{journal}{Naturw.} \textbf{\bibinfo{volume}{45}},
  \bibinfo{pages}{538} (\bibinfo{year}{1958}{\natexlab{b}}).

\bibitem[{\citenamefont{Hannon and Trammell}(1999)}]{Hannon:1999fk}
\bibinfo{author}{\bibfnamefont{J.~P.} \bibnamefont{Hannon}} \bibnamefont{and}
  \bibinfo{author}{\bibfnamefont{G.~T.} \bibnamefont{Trammell}},
  \bibinfo{journal}{Hyp. Int.} \textbf{\bibinfo{volume}{123-124}},
  \bibinfo{pages}{127} (\bibinfo{year}{1999}).

\bibitem[{\citenamefont{Smirnov et~al.}(2005)\citenamefont{Smirnov, van
  B{\"u}rck, Potzel, Schindelmann, Popov, Gerdau, Shvyd'ko, R{\"u}ter, and
  Leupold}}]{Smirnov:2005uq}
\bibinfo{author}{\bibfnamefont{G.~V.} \bibnamefont{Smirnov}},
  \bibinfo{author}{\bibfnamefont{U.}~\bibnamefont{van B{\"u}rck}},
  \bibinfo{author}{\bibfnamefont{W.}~\bibnamefont{Potzel}},
  \bibinfo{author}{\bibfnamefont{P.}~\bibnamefont{Schindelmann}},
  \bibinfo{author}{\bibfnamefont{S.~L.} \bibnamefont{Popov}},
  \bibinfo{author}{\bibfnamefont{E.}~\bibnamefont{Gerdau}},
  \bibinfo{author}{\bibfnamefont{Y.~V.} \bibnamefont{Shvyd'ko}},
  \bibinfo{author}{\bibfnamefont{H.~D.} \bibnamefont{R{\"u}ter}},
  \bibnamefont{and} \bibinfo{author}{\bibfnamefont{O.}~\bibnamefont{Leupold}},
  \bibinfo{journal}{Phys. Rev. A} \textbf{\bibinfo{volume}{71}},
  \bibinfo{pages}{023804} (\bibinfo{year}{2005}).

\bibitem[{\citenamefont{Habs et~al.}(2009)\citenamefont{Habs, Tajima,
  Schreiber, Barty, Fujiwara, and Thirolf}}]{Habs:2009uq}
\bibinfo{author}{\bibfnamefont{D.}~\bibnamefont{Habs}},
  \bibinfo{author}{\bibfnamefont{T.}~\bibnamefont{Tajima}},
  \bibinfo{author}{\bibfnamefont{J.}~\bibnamefont{Schreiber}},
  \bibinfo{author}{\bibfnamefont{C.~P.~J.} \bibnamefont{Barty}},
  \bibinfo{author}{\bibfnamefont{M.}~\bibnamefont{Fujiwara}}, \bibnamefont{and}
  \bibinfo{author}{\bibfnamefont{P.~G.} \bibnamefont{Thirolf}},
  \bibinfo{journal}{Eur. Phys. J. D} \textbf{\bibinfo{volume}{55}},
  \bibinfo{pages}{279} (\bibinfo{year}{2009}).

\bibitem[{\citenamefont{Dicke}(1954)}]{Dicke:1954kx}
\bibinfo{author}{\bibfnamefont{R.~H.} \bibnamefont{Dicke}},
  \bibinfo{journal}{Phys. Rev.} \textbf{\bibinfo{volume}{93}},
  \bibinfo{pages}{99} (\bibinfo{year}{1954}).

\bibitem[{\citenamefont{Scully et~al.}(2006)\citenamefont{Scully, Fry, Ooi, and
  W\'odkiewicz}}]{Scully:2006fk}
\bibinfo{author}{\bibfnamefont{M.~O.} \bibnamefont{Scully}},
  \bibinfo{author}{\bibfnamefont{E.~S.} \bibnamefont{Fry}},
  \bibinfo{author}{\bibfnamefont{C.~H.~R.} \bibnamefont{Ooi}},
  \bibnamefont{and}
  \bibinfo{author}{\bibfnamefont{K.}~\bibnamefont{W\'odkiewicz}},
  \bibinfo{journal}{Phys. Rev. Lett.} \textbf{\bibinfo{volume}{96}},
  \bibinfo{pages}{010501} (\bibinfo{year}{2006}).

\bibitem[{\citenamefont{Scully}(2007)}]{Scully:2007fk}
\bibinfo{author}{\bibfnamefont{M.~O.} \bibnamefont{Scully}},
  \bibinfo{journal}{Laser Phys.} \textbf{\bibinfo{volume}{17}},
  \bibinfo{pages}{635} (\bibinfo{year}{2007}).

\bibitem[{\citenamefont{Sete et~al.}(2010)\citenamefont{Sete, Svidzinsky,
  Eleuch, Yang, Nevels, and Scully}}]{Sete:2010fu}
\bibinfo{author}{\bibfnamefont{E.~A.} \bibnamefont{Sete}},
  \bibinfo{author}{\bibfnamefont{A.~A.} \bibnamefont{Svidzinsky}},
  \bibinfo{author}{\bibfnamefont{H.}~\bibnamefont{Eleuch}},
  \bibinfo{author}{\bibfnamefont{Z.}~\bibnamefont{Yang}},
  \bibinfo{author}{\bibfnamefont{R.~D.} \bibnamefont{Nevels}},
  \bibnamefont{and} \bibinfo{author}{\bibfnamefont{M.~O.}
  \bibnamefont{Scully}}, \bibinfo{journal}{J. Mod. Opt.}
  \textbf{\bibinfo{volume}{57}}, \bibinfo{pages}{1311} (\bibinfo{year}{2010}).

\bibitem[{\citenamefont{Burnham and Chiao}(1969)}]{Burnham:1969uq}
\bibinfo{author}{\bibfnamefont{D.~C.} \bibnamefont{Burnham}} \bibnamefont{and}
  \bibinfo{author}{\bibfnamefont{R.~Y.} \bibnamefont{Chiao}},
  \bibinfo{journal}{Phys. Rev.} \textbf{\bibinfo{volume}{188}},
  \bibinfo{pages}{667} (\bibinfo{year}{1969}).

\bibitem[{\citenamefont{Fr{\"o}hlich et~al.}(1991)\citenamefont{Fr{\"o}hlich,
  Kulik, Uebbing, Mysyrowicz, Langer, Stolz, and von~der
  Osten}}]{Frohlich:1991kx}
\bibinfo{author}{\bibfnamefont{D.}~\bibnamefont{Fr{\"o}hlich}},
  \bibinfo{author}{\bibfnamefont{A.}~\bibnamefont{Kulik}},
  \bibinfo{author}{\bibfnamefont{B.}~\bibnamefont{Uebbing}},
  \bibinfo{author}{\bibfnamefont{A.}~\bibnamefont{Mysyrowicz}},
  \bibinfo{author}{\bibfnamefont{V.}~\bibnamefont{Langer}},
  \bibinfo{author}{\bibfnamefont{H.}~\bibnamefont{Stolz}}, \bibnamefont{and}
  \bibinfo{author}{\bibfnamefont{W.}~\bibnamefont{von~der Osten}},
  \bibinfo{journal}{Phys. Rev. Lett.} \textbf{\bibinfo{volume}{67}},
  \bibinfo{pages}{2343} (\bibinfo{year}{1991}).

\bibitem[{\citenamefont{van B{\"u}rck}(1999)}]{Burck:1999fk}
\bibinfo{author}{\bibfnamefont{U.}~\bibnamefont{van B{\"u}rck}},
  \bibinfo{journal}{Hyp. Int.} \textbf{\bibinfo{volume}{123-124}},
  \bibinfo{pages}{483} (\bibinfo{year}{1999}).

\bibitem[{\citenamefont{R{\"o}hlsberger
  et~al.}(2010)\citenamefont{R{\"o}hlsberger, Schlage, Sahoo, Couet, and
  R{\"u}ffer}}]{Rohlsberger:2010fk}
\bibinfo{author}{\bibfnamefont{R.}~\bibnamefont{R{\"o}hlsberger}},
  \bibinfo{author}{\bibfnamefont{K.}~\bibnamefont{Schlage}},
  \bibinfo{author}{\bibfnamefont{B.}~\bibnamefont{Sahoo}},
  \bibinfo{author}{\bibfnamefont{S.}~\bibnamefont{Couet}}, \bibnamefont{and}
  \bibinfo{author}{\bibfnamefont{R.}~\bibnamefont{R{\"u}ffer}},
  \bibinfo{journal}{Science} \textbf{\bibinfo{volume}{328}},
  \bibinfo{pages}{1248} (\bibinfo{year}{2010}).

\bibitem[{\citenamefont{P{\'a}lffy et~al.}(2011)\citenamefont{P{\'a}lffy,
  Keitel, and Evers}}]{Palffy:2011vn}
\bibinfo{author}{\bibfnamefont{A.}~\bibnamefont{P{\'a}lffy}},
  \bibinfo{author}{\bibfnamefont{C.~H.} \bibnamefont{Keitel}},
  \bibnamefont{and} \bibinfo{author}{\bibfnamefont{J.}~\bibnamefont{Evers}},
  \bibinfo{journal}{Phys. Rev. B} \textbf{\bibinfo{volume}{83}},
  \bibinfo{pages}{155103} (\bibinfo{year}{2011}).

\bibitem[{\citenamefont{Rohlsberger et~al.}(2012)\citenamefont{Rohlsberger,
  Wille, Schlage, and Sahoo}}]{Rohlsberger:2012uq}
\bibinfo{author}{\bibfnamefont{R.}~\bibnamefont{Rohlsberger}},
  \bibinfo{author}{\bibfnamefont{H.-C.} \bibnamefont{Wille}},
  \bibinfo{author}{\bibfnamefont{K.}~\bibnamefont{Schlage}}, \bibnamefont{and}
  \bibinfo{author}{\bibfnamefont{B.}~\bibnamefont{Sahoo}},
  \bibinfo{journal}{Nature} \textbf{\bibinfo{volume}{482}},
  \bibinfo{pages}{199} (\bibinfo{year}{2012}).

\bibitem[{\citenamefont{Sch{\"u}tzhold et~al.}()\citenamefont{Sch{\"u}tzhold,
  Habs, Thirolf, and Fujiwara}}]{ELIwhitebook}
\bibinfo{author}{\bibfnamefont{R.}~\bibnamefont{Sch{\"u}tzhold}},
  \bibinfo{author}{\bibfnamefont{D.}~\bibnamefont{Habs}},
  \bibinfo{author}{\bibfnamefont{P.}~\bibnamefont{Thirolf}}, \bibnamefont{and}
  \bibinfo{author}{\bibfnamefont{M.}~\bibnamefont{Fujiwara}},
  \emph{\bibinfo{title}{{{\rm in} The White Book of ELI Nuclear Physics, {\tt
  http://www.eli-np.ro/}, {\rm pp. 114--115}}}}.

\bibitem[{\citenamefont{Lipkin}(2002)}]{Lipkin:2002fk}
\bibinfo{author}{\bibfnamefont{H.~J.} \bibnamefont{Lipkin}},
  \emph{\bibinfo{title}{{{\rm in} Multiple Facets of Quantization and
  Supersymmetry, {\rm edited by M. Olshanetsky and A. Vainshtein}}}}, pp.
  128--150 (\bibinfo{publisher}{World Scientific},
  \bibinfo{address}{Singapore}, \bibinfo{year}{2002}).

\bibitem[{\citenamefont{Rehler and Eberly}(1971)}]{Rehler:1971fk}
\bibinfo{author}{\bibfnamefont{N.~E.} \bibnamefont{Rehler}} \bibnamefont{and}
  \bibinfo{author}{\bibfnamefont{J.~H.} \bibnamefont{Eberly}},
  \bibinfo{journal}{Phys. Rev. A} \textbf{\bibinfo{volume}{3}},
  \bibinfo{pages}{1735} (\bibinfo{year}{1971}).

\bibitem[{\citenamefont{Scully and Zubairy}(1997)}]{Scully:1997fk}
\bibinfo{author}{\bibfnamefont{M.~O.} \bibnamefont{Scully}} \bibnamefont{and}
  \bibinfo{author}{\bibfnamefont{M.~S.} \bibnamefont{Zubairy}},
  \emph{\bibinfo{title}{{Quantum Optics}}} (\bibinfo{publisher}{Cambridge
  University Press}, \bibinfo{address}{Cambridge}, \bibinfo{year}{1997}).

\bibitem[{\citenamefont{Junker et~al.}()\citenamefont{Junker, P{\'a}lffy, and
  Keitel}}]{Junker:2012fk}
\bibinfo{author}{\bibfnamefont{A.}~\bibnamefont{Junker}},
  \bibinfo{author}{\bibfnamefont{A.}~\bibnamefont{P{\'a}lffy}},
  \bibnamefont{and} \bibinfo{author}{\bibfnamefont{C.~H.}
  \bibnamefont{Keitel}}, \bibinfo{note}{e-print arXiv:1203.2149}.

\bibitem[{\citenamefont{Shvyd'ko et~al.}(1996)\citenamefont{Shvyd'ko, Hertrich,
  van B{\"u}rck, Gerdau, Leupold, Metge, R{\"u}ter, Schwendy, Smirnov, Potzel
  et~al.}}]{Shvydko:1996fk}
\bibinfo{author}{\bibfnamefont{Y.~V.} \bibnamefont{Shvyd'ko}},
  \bibinfo{author}{\bibfnamefont{T.}~\bibnamefont{Hertrich}},
  \bibinfo{author}{\bibfnamefont{U.}~\bibnamefont{van B{\"u}rck}},
  \bibinfo{author}{\bibfnamefont{E.}~\bibnamefont{Gerdau}},
  \bibinfo{author}{\bibfnamefont{O.}~\bibnamefont{Leupold}},
  \bibinfo{author}{\bibfnamefont{J.}~\bibnamefont{Metge}},
  \bibinfo{author}{\bibfnamefont{H.~D.} \bibnamefont{R{\"u}ter}},
  \bibinfo{author}{\bibfnamefont{S.}~\bibnamefont{Schwendy}},
  \bibinfo{author}{\bibfnamefont{G.~V.} \bibnamefont{Smirnov}},
  \bibinfo{author}{\bibfnamefont{W.}~\bibnamefont{Potzel}},
  \bibnamefont{et~al.}, \bibinfo{journal}{Phys. Rev. Lett.}
  \textbf{\bibinfo{volume}{77}}, \bibinfo{pages}{3232} (\bibinfo{year}{1996}).

\bibitem[{\citenamefont{R{\"o}hlsberger
  et~al.}(2000)\citenamefont{R{\"o}hlsberger, Toellner, Sturhahn, Quast, Alp,
  Bernhard, Burkel, Leupold, and Gerdau}}]{Rohlsberger:2000fk}
\bibinfo{author}{\bibfnamefont{R.}~\bibnamefont{R{\"o}hlsberger}},
  \bibinfo{author}{\bibfnamefont{T.~S.} \bibnamefont{Toellner}},
  \bibinfo{author}{\bibfnamefont{W.}~\bibnamefont{Sturhahn}},
  \bibinfo{author}{\bibfnamefont{K.~W.} \bibnamefont{Quast}},
  \bibinfo{author}{\bibfnamefont{E.~E.} \bibnamefont{Alp}},
  \bibinfo{author}{\bibfnamefont{A.}~\bibnamefont{Bernhard}},
  \bibinfo{author}{\bibfnamefont{E.}~\bibnamefont{Burkel}},
  \bibinfo{author}{\bibfnamefont{O.}~\bibnamefont{Leupold}}, \bibnamefont{and}
  \bibinfo{author}{\bibfnamefont{E.}~\bibnamefont{Gerdau}},
  \bibinfo{journal}{Phys. Rev. Lett.} \textbf{\bibinfo{volume}{84}},
  \bibinfo{pages}{1007} (\bibinfo{year}{2000}).

\bibitem[{\citenamefont{Coussement et~al.}(2002)\citenamefont{Coussement,
  Rostovtsev, Odeurs, Neyens, Muramatsu, Gheysen, Callens, Vyvey, Kozyreff,
  Mandel et~al.}}]{Coussement:2002zr}
\bibinfo{author}{\bibfnamefont{R.}~\bibnamefont{Coussement}},
  \bibinfo{author}{\bibfnamefont{Y.}~\bibnamefont{Rostovtsev}},
  \bibinfo{author}{\bibfnamefont{J.}~\bibnamefont{Odeurs}},
  \bibinfo{author}{\bibfnamefont{G.}~\bibnamefont{Neyens}},
  \bibinfo{author}{\bibfnamefont{H.}~\bibnamefont{Muramatsu}},
  \bibinfo{author}{\bibfnamefont{S.}~\bibnamefont{Gheysen}},
  \bibinfo{author}{\bibfnamefont{R.}~\bibnamefont{Callens}},
  \bibinfo{author}{\bibfnamefont{K.}~\bibnamefont{Vyvey}},
  \bibinfo{author}{\bibfnamefont{G.}~\bibnamefont{Kozyreff}},
  \bibinfo{author}{\bibfnamefont{P.}~\bibnamefont{Mandel}},
  \bibnamefont{et~al.}, \bibinfo{journal}{Phys. Rev. Lett.}
  \textbf{\bibinfo{volume}{89}}, \bibinfo{pages}{107601}
  (\bibinfo{year}{2002}).

\bibitem[{\citenamefont{P{\'a}lffy et~al.}(2009)\citenamefont{P{\'a}lffy,
  Keitel, and Evers}}]{Palffy:2009kx}
\bibinfo{author}{\bibfnamefont{A.}~\bibnamefont{P{\'a}lffy}},
  \bibinfo{author}{\bibfnamefont{C.~H.} \bibnamefont{Keitel}},
  \bibnamefont{and} \bibinfo{author}{\bibfnamefont{J.}~\bibnamefont{Evers}},
  \bibinfo{journal}{Phys. Rev. Lett.} \textbf{\bibinfo{volume}{103}},
  \bibinfo{pages}{017401} (\bibinfo{year}{2009}).

\bibitem[{\citenamefont{Adams}(2011)}]{Adams:2011ys}
\bibinfo{author}{\bibfnamefont{B.~W.} \bibnamefont{Adams}},
  \bibinfo{journal}{J. Mod. Opt.} \textbf{\bibinfo{volume}{58}},
  \bibinfo{pages}{1638} (\bibinfo{year}{2011}).

\bibitem[{\citenamefont{Kim et~al.}(2008)\citenamefont{Kim, Shvyd'ko, and
  Reiche}}]{Kim:2008qf}
\bibinfo{author}{\bibfnamefont{K.-J.} \bibnamefont{Kim}},
  \bibinfo{author}{\bibfnamefont{Y.}~\bibnamefont{Shvyd'ko}}, \bibnamefont{and}
  \bibinfo{author}{\bibfnamefont{S.}~\bibnamefont{Reiche}},
  \bibinfo{journal}{Phys. Rev. Lett.} \textbf{\bibinfo{volume}{100}},
  \bibinfo{pages}{244802} (\bibinfo{year}{2008}).

\bibitem[{\citenamefont{B{\"u}rvenich et~al.}(2006)\citenamefont{B{\"u}rvenich,
  Evers, and Keitel}}]{Burvenich:2006bh}
\bibinfo{author}{\bibfnamefont{T.~J.} \bibnamefont{B{\"u}rvenich}},
  \bibinfo{author}{\bibfnamefont{J.}~\bibnamefont{Evers}}, \bibnamefont{and}
  \bibinfo{author}{\bibfnamefont{C.~H.} \bibnamefont{Keitel}},
  \bibinfo{journal}{Phys. Rev. Lett.} \textbf{\bibinfo{volume}{96}},
  \bibinfo{pages}{142501} (\bibinfo{year}{2006}).

\bibitem[{\citenamefont{P{\'a}lffy et~al.}(2008)\citenamefont{P{\'a}lffy,
  Evers, and Keitel}}]{Palffy:2008uq}
\bibinfo{author}{\bibfnamefont{A.}~\bibnamefont{P{\'a}lffy}},
  \bibinfo{author}{\bibfnamefont{J.}~\bibnamefont{Evers}}, \bibnamefont{and}
  \bibinfo{author}{\bibfnamefont{C.~H.} \bibnamefont{Keitel}},
  \bibinfo{journal}{Phys. Rev. C} \textbf{\bibinfo{volume}{77}},
  \bibinfo{pages}{044602} (\bibinfo{year}{2008}).

\bibitem[{\citenamefont{Liao et~al.}(2011)\citenamefont{Liao, P{\'a}lffy, and
  Keitel}}]{Liao:2011cr}
\bibinfo{author}{\bibfnamefont{W.-T.} \bibnamefont{Liao}},
  \bibinfo{author}{\bibfnamefont{A.}~\bibnamefont{P{\'a}lffy}},
  \bibnamefont{and} \bibinfo{author}{\bibfnamefont{C.~H.}
  \bibnamefont{Keitel}}, \bibinfo{journal}{Phys. Lett. B}
  \textbf{\bibinfo{volume}{705}}, \bibinfo{pages}{134} (\bibinfo{year}{2011}).

\bibitem[{fel()}]{felbasics}
\emph{\bibinfo{title}{{\tt
  http://hasylab.desy.de/e70/e6129/e4242/e4370/\\felbasics\_eng.pdf}}}.

\bibitem[{\citenamefont{Siwick et~al.}(2003)\citenamefont{Siwick, Dwyer,
  Jordan, and Miller}}]{Siwick:2003fk}
\bibinfo{author}{\bibfnamefont{B.~J.} \bibnamefont{Siwick}},
  \bibinfo{author}{\bibfnamefont{J.~R.} \bibnamefont{Dwyer}},
  \bibinfo{author}{\bibfnamefont{R.~E.} \bibnamefont{Jordan}},
  \bibnamefont{and} \bibinfo{author}{\bibfnamefont{R.~J.~D.}
  \bibnamefont{Miller}}, \bibinfo{journal}{Science}
  \textbf{\bibinfo{volume}{302}}, \bibinfo{pages}{1382} (\bibinfo{year}{2003}).

\bibitem[{\citenamefont{Holstein and Primakoff}(1940)}]{Holstein:1940vn}
\bibinfo{author}{\bibfnamefont{T.}~\bibnamefont{Holstein}} \bibnamefont{and}
  \bibinfo{author}{\bibfnamefont{H.}~\bibnamefont{Primakoff}},
  \bibinfo{journal}{Phys. Rev.} \textbf{\bibinfo{volume}{58}},
  \bibinfo{pages}{1098} (\bibinfo{year}{1940}).

\end{thebibliography}

\end{document}